# Exploiting dynamic nonlinearity in upconversion nanoparticles for super-resolution imaging

Chaohao Chen[1,2†], Lei Ding[3†], Baolei Liu[4], Ziqin Du[3], Yongtao Liu[5], Xiangjun Di[3], Xuchen Shan[4], Chenxiao Lin[6], Min Zhang[1], Xiaoxue Xu[7], Xiaolan Zhong[4], Jianfeng Wang[4], Lingqian Chang[8], Ben J. Halkon[9], Xin Chen[2], Faliang Cheng[1*], Fan Wang[4,10*]

[1]Guangdong Engineering and Technology Research Center for Advanced Nanomaterials, School of Environment and Civil Engineering, Dongguan University of Technology, Dongguan 523808, China

[2]Department of Chemical Engineering, Shaanxi Key Laboratory of Energy Chemical Process Intensification, Institute of Polymer Science in Chemical Engineering, School of Chemical Engineering and Technology, Xi'an Jiaotong University, Xi'an 710049, China

[3]School of Mathematical and Physical Sciences, Faculty of Science, University of Technology Sydney, NSW 2007, Australia

[4]School of Physics, Beihang University, Beijing 100191, China

[5]Smart Computational Imaging Laboratory, Nanjing University of Science and Technology, Nanjing, Jiangsu Province 210094, China

[6]Department for Electrochemical Energy Storage, Helmholtz-Zentrum Berlin für Materialien und Energie, Hahn-Meitner-Platz, Berlin 14109, Germany

[7]School of Biomedical Engineering, Faculty of Engineering and Information Technology, University of Technology Sydney, NSW 2007, Australia

[8]Beijing Advanced Innovation Center for Biomedical Engineering, School of Biological Science and Medical Engineering, Beihang University, Beijing 100191, China

[9]Centre for Audio, Acoustics & Vibration, Faculty of Engineering & IT, University of Technology Sydney, Ultimo, NSW 2007, Australia

[10]School of Electrical and Data Engineering, Faculty of Engineering and Information Technology, University of Technology Sydney, NSW 2007, Australia.

[†]These authors contributed equally: Chaohao Chen, Lei Ding

[*]Correspondence to fan.wang@uts.edu.au, chengfl@dgut.edu.cn.

**Abstract**

**Single-beam super-resolution microscopy, also known as superlinear microscopy, exploits the nonlinear response of fluorescent probes in confocal microscopy. The technique requires no complex purpose-built system, light field modulation, or beam shaping. Here, we present a strategy to enhance this technique's spatial resolution by modulating excitation intensity during image acquisition. This modulation induces dynamic optical nonlinearity in upconversion nanoparticles (UCNPs), resulting in variations of higher spatial-frequency information in the obtained images. The high-order information can be extracted with a proposed weighted finite difference imaging algorithm from raw fluorescence images, to generate an image with a higher resolution than superlinear microscopy images. We apply this approach to resolve two adjacent nanoparticles within a diffraction-limited area, improving the**

resolution to 130 nm. This work suggests a new scope for developing dynamic nonlinear fluorescent probes in super-resolution nanoscopy.

## Introduction

Optical super-resolution techniques that break through the limits of diffraction to achieve high resolution are a key source of advances in life sciences. The fundamental principle behind super-resolution techniques is to obtain sub-diffraction image details, also known as high spatial-frequency information.[1] In super-resolution microscopes, the nonlinear technique is a widely used method by employing strong nonlinearity in the excitation and emission rates of fluorescent probes. Nonlinear techniques induce patterned light excitation on wide-field microscopes, such as fringes[2,3] and speckles[4,5], that can get higher spatial-frequency information via Fourier transformation. Saturation excitation microscopy[6] employs harmonically controlled excitation intensity for point scanning sub-diffraction imaging. The resulting smaller effective point spread function (PSF) enhances the spatial resolution. However, both these techniques require complicated illumination modules to fulfill the excitation conditions required for acquiring the hidden higher spatial-frequency information.

Superlinear emission[7] achieves super-resolution imaging on conventional confocal microscopy in a nonlinear modality that does not need sophisticated modulation. Superlinear emission can generate steep spatial modulations that are narrower than the excitation beam during the scanning process.[8] Nanodiamond possesses a superlinear operation state under multiple lasers and specialized illumination sequences.[9] Quantum dots' superlinear response also has potential for sub-diffraction imaging applications.[10] However, the need for complicated setups or high excitation powers complicates the superlinear emission technique.

Upconversion nanoparticles (UCNPs)[11–14] are one of the most promising nonlinear probes. They can achieve efficient superlinearity without sophisticated techniques or high excitation power.[15–19] UCNPs are composed of a nanocrystal host and thousands of embedded lanthanide ions such as $Yb^{3+}$, $Nd^{3+}$, $Tm^{3+}$, and $Er^{3+}$. The lanthanide ions contain multiple and long-lived intermediate energy states that facilitate energy transfers, resulting in upconversion emissions ranging from near-infrared (NIR) to visible and ultraviolet.[20–24] Due to their unique optical characteristics, UCNPs have been widely used in super-resolution imaging applications.[25–34] Recently, the superlinear response of a single UCNP from the photon avalanche effect enabled sub-70 nm spatial resolution, but at the expense of imaging speed.[35,36]

This work takes advantage of the dynamic nonlinearity of UCNPs to improve the achievable super-resolution of the superlinear imaging modality. We demonstrate that superlinearity in UCNP is dynamically tunable by varying excitation intensity during image acquisition. This dynamic nonlinearity can encode variations of spatial-frequency information in the obtained images. We then show the ability to extract higher spatial-frequency information through a weighted finite difference method. We examine the approach through both numerical simulations and imaging experiments, resolving two adjacent nanoparticles at a sub-diffraction distance.

**Results and discussion**

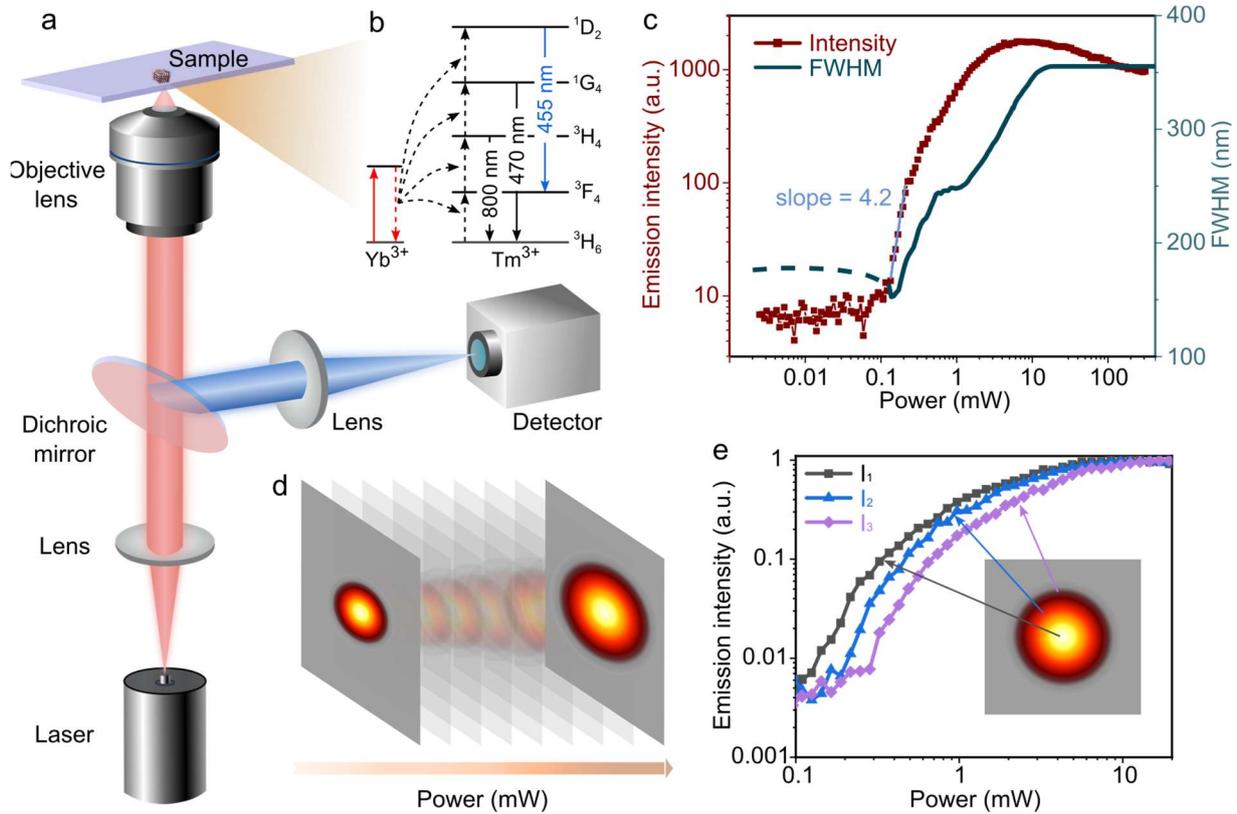

**Figure 1. Dynamic nonlinearity in UCNPs**. (a) Schematic diagram of the confocal setup. A single-photon counting avalanche photodiode (SPAD) captures fluorescence intensities under differential excitation powers. (b) The simplified energy level and upconversion process of the nanoparticles NaYF$_4$: 40% Yb$^{3+}$, 4% Tm$^{3+}$ under 980 nm continuous-wave excitation. The steep emission nonlinearity in UCNPs mainly arises from the four-photon excited state (455 nm, $^1D_2 \to$ $^3F_4$). (c) The measured nonlinear saturation intensities and their corresponding simulated full width at half maximum (FWHM) of a single nanoparticle at 455 nm emissions function of excitation intensity. (d) Schematic representation of the acquired confocal images of a single nanoparticles' 455 nm emission band under the gradient of excitation intensities. (e) The corresponding saturation curves at different positions in the PSF. The excitation intensity is measured at the back aperture of the objective lens.

To demonstrate the feature of dynamic nonlinearity in a single lanthanide-doped nanoparticle, we employ a NaYF$_4$ nanocrystal host co-doped with high concentrations of 40% Yb$^{3+}$ sensitizer ions and 4% Tm$^{3+}$ activators ions (41 nm average size, Figure S1). We conduct the experiments and data acquisitions under differential excitation intensities in a conventional confocal configuration with 980 nm continuous-wave lasers (Figure 1a). The multiple and long-lived energy levels (Figure 1b) of UCNPs can produce various multiphoton nonlinear upconversion emissions. The energy transfer processes include cross-relaxation between activator ions and excited-state absorption.[31,32] This work uses four-photon 455 nm emission ($^1D_2 \to$ $^3F_4$) for super-resolution imaging because of the sharp slope in the optical nonlinear response curve (Figure 1c). When the excitation power is less than 0.1 mW, the weak emission signal is buried in the noise and background. Increasing the excitation power to 0.1 mW exhibits a substantial nonlinear optical response with a sudden rise in emission intensity. However, an ever-increasing excitation power will affect carrier generation and upconversion rates. The saturation point occurs when the distribution in this state establishes a dynamic equilibrium between intermediate states. The slope gradient decreases as the excitation intensity increases to 100 mW due to the dynamic saturation of photon carriers.

The large nonlinearity in UCNPs has previously achieved sub-diffraction imaging in conventional confocal configurations.[8] The strong nonlinear emissions occur in the central region of the PSF, which is smaller than the size of the excitation beam. According to the superlinear imaging equation: $\Delta r = \frac{\lambda}{2NA\sqrt{N}}$, the emission PSF scales inversely with the square root of the order of nonlinearity.[7] Due to the dynamic nonlinearity of UCNPs, the full width at half maximum (FWHM) of the emission PSFs is adjustable as a function of the excitation intensities, as shown in Figure 1c. With a given nonlinear saturation intensities curve, the resolution can be optimized at 190 nm.[8] The emission PSFs progressively become larger with the increased excitation intensity (Figure 1d), demonstrating stepwise optical saturation. Figure 1e shows three saturation curves at different points in the PSF. Compared with the peripheral points, the central point occurs the sharp nonlinear response and saturation phenomena first.

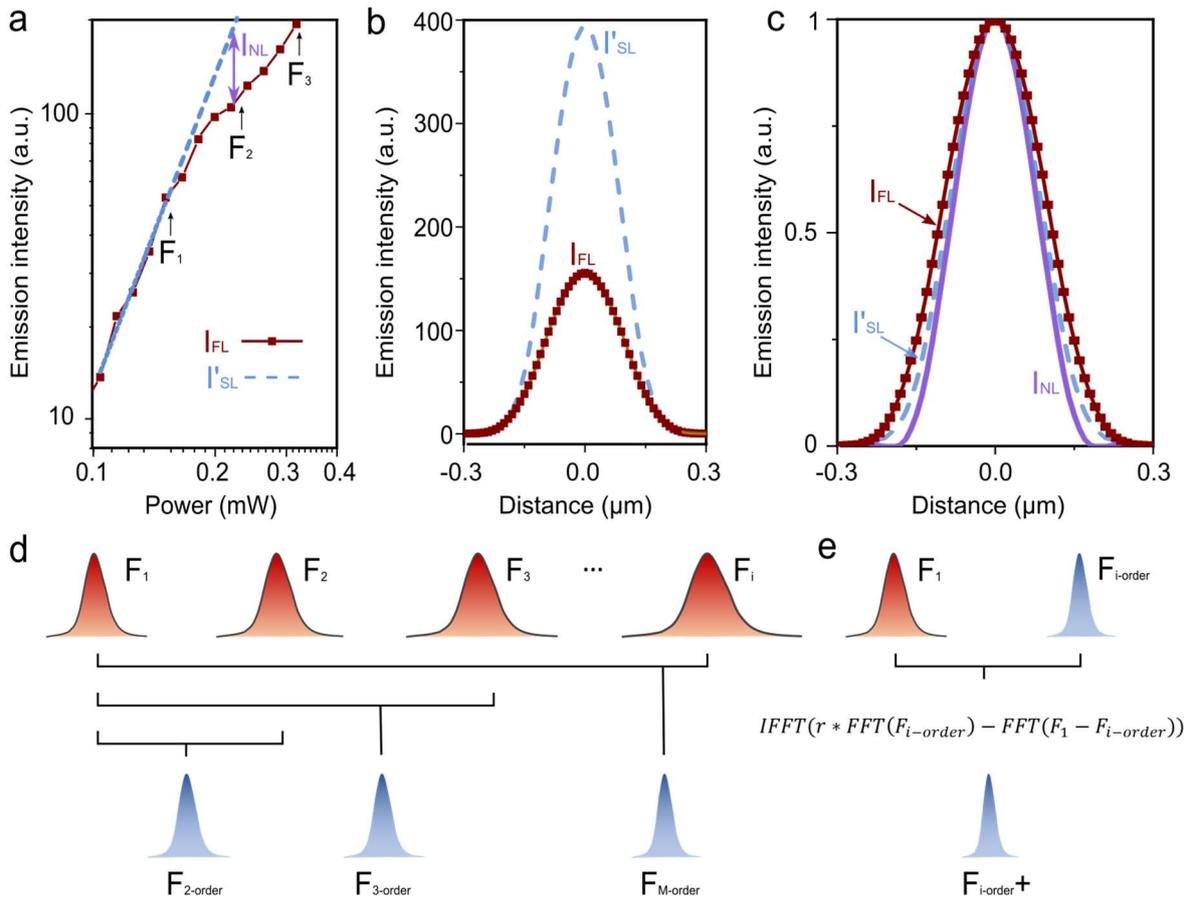

**Figure 2. Extracting the higher spatial-frequency information to improve resolution.** (a) Relationship between emission and excitation intensity. Fluorescent intensity ($I_{FL}$), ideal superlinear response intensity ($I'_{SL}$), and differential nonlinear component ($I_{NL}$). (b) Comparison of estimated superlinear response intensity and fluorescent intensity distribution in the PSF under the same excitation power. (c) Extracting the higher spatial-frequency information to provide a narrower PSF. (d) Concept of improving the spatial resolution with weighted finite difference method from two to M steps. (e) Concept of the Fourier modulation process to further enhance the resolution.

Figure 2a shows the measured 455 nm fluorescence intensity ($I_{FL}$) at the superlinear area of Figure 1c, revealing that the emission signal is nonlinearly proportional to the excitation intensity. Notably, a weak saturation phenomenon of the upconversion emission can occur when the excitation power exceeds 0.15 mW. If there is no saturation effect, the optical response should correspond to the dashed line, which represents the ideal superlinear response intensity ($I'_{SL}$). The difference ($I_{NL}$) between $I_{FL}$ and $I'_{SL}$ contains a high amplitude of the higher-order power-dependent information, which leads to a dramatic change with the increase of excitation power. The $I_{NL}$ becomes more noticeable when a Gaussian excitation profile is applied (Figure 2b), where the $I_{NL}$ become larger with an excitation location closer to the central region. The intensity change with distance in $I_{NL}$ is even large than that for $I_{FL}$ or $I'_{SL}$.

As a consequence, a differential operation between the fluorescence signals at various excitation powers can extract the higher spatial-frequency information that contributes to the resolution improvement (Figure 2c).

We next theoretically demonstrate the proposed concept to achieve the resolution enhancement by using the dynamic nonlinearity of UCNPs. We simplify the dual-doped $Yb^{3+}$-$Tm^{3+}$ ions system into a six-energy-level model (Figure S3 and Table S1) and build rate equations (Supporting Information). By neglecting higher excited states and photobleaching, the model reproduces the experimentally observed superlinear emission behaviors (Figure S4). Although the overall 455 nm emission band response curve involves various energy transfer mechanisms, we find that the emission at low excitation power is similar to a conventional multiphoton phenomenon. Due to the superlinear emission property of UCNPs, the low excitation intensity minimizes resolution loss and allows us to undertake Taylor series analysis. Following a similarly reported derivation[37], we can express the fluorescence intensities of the $i$-th step $F_i(x)$ among the $M$ images, as a sum of linear and nonlinear components.

$$F_i(x) = KN_0(aI_{0i}^N g^N(x) - a^2 I_{0i}^{2N} g^{2N}(x) + a^3 I_{0i}^{3N} g^{3N}(x) - \cdots) \qquad (1)$$

where $K = \psi_F/\tau$, $\psi_F$ is the fluorescence detection efficiency, $\tau$ is the fluorescence lifetime, $N_0$ is the concentration of the fluorophore, $a = \tau\sigma_N(\lambda/hc)^N$, $N$ is $N$-photon excitation, $\sigma_N$ is the cross-section for $N$-photon excitation, $\lambda$ is the excitation wavelength, $h$ is Planck's constant, $c$ is the velocity of light, $I_{0i}$ is the focal irradiance of the i-th step, $g(x) = \exp(-2x^2/\omega_0^2)$ is the Gaussian profile of excitation spot, $\omega_0$ is the diffraction limit.

For $N$-photon excitation, the profile $g(x)$ is exponent up to the $N$-th, as $g^N(x) = \exp(-2x^2/(\omega_0/\sqrt{N})^2)$. Thus, we can have a $\sqrt{N}$ Improvement of the resolution for $N$-photon superlinear microscopy.[7] Moreover, the higher exponent of $g^N(x)$, such as $g^{2N}(x)$, $g^{3N}(x)$, indicates components of the higher spatial-frequency information. According to Equation (1), the lowest exponent of $g^N(x)$ dominates the spatial resolution of $F_i(x)$, extracting higher exponents of $g^N(x)$ would increase the resolution. We can eliminate the 1 to $M$-1 exponents of $g^N(x)$ by combining $M$ steps of confocal images with a weighted finite difference method, giving the lowest exponent as $g^{MN}(x)$, generating an image with $\sqrt{M}$-fold higher resolution than $N$-photon superlinear microscopy (Figure 2d, see detail in Supporting information), and total enhancement is $\sqrt{MN}$-fold than the diffraction limit. For example, we can obtain a $\sqrt{2}$-fold increase in spatial resolution by combining two images. Table S2 displays the established weighted coefficients for the image processing. We employ Fourier modulation[38] based on the results from a weighted finite difference method to extract higher spatial-frequency information and increase resolution (Figure 2e, see detail in Supporting information). It is worth noting that the more photons involved in the fluorescence generation, the greater variation in its nonlinear response slope, which increases resolution in the final computational step.

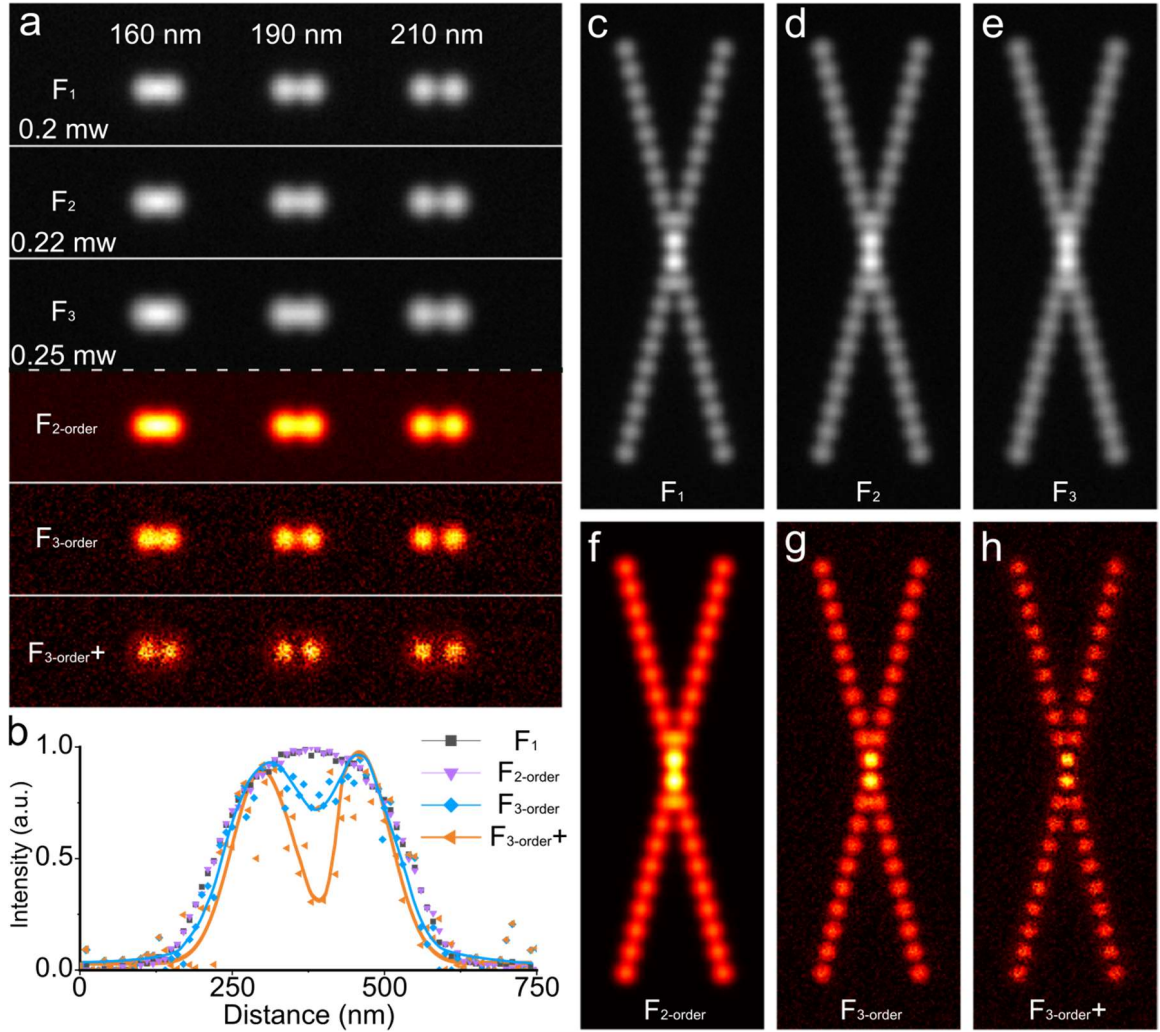

**Figure 3. Simulation for resolving adjacent nanoparticles and X-bars.** (a) Simulated confocal imaging (gray) of two adjacent emitters 160 nm, 190 nm, and 210 nm apart under different excitation intensities of 0.2 mW, 0.22 mW, and 0.25 mW. Super-resolution imaging (hot) results from the second or third-order weighted finite difference method for comparison. The $F_{3\text{-order}}+$ represents the Fourier modulation process based on the results from a third-order weighted finite difference method. (b) The cross-section profiles of the simulated images of the nanoparticles with 160 nm distance in (a). (c-h) The simulation of the X-bar cross line structure with sparse UCNPs.

Next, we conduct an imaging simulation using the above differential excitation technique to extract higher spatial-frequency information from superlinear images (Figure 3a). The fluorescent confocal images of two emitters 160 nm, 190 nm, and 210 nm apart are simulated under excitation powers of 0.2 mW, 0.22 mW, and 0.25 mW. The fluorescence wavelength is 455 nm, as same as the four-photon emission band in the UCNPs. We input the complete set of simulated images with the superlinear approach and weighted finite difference method algorithms for comparison. From the cross-section profiles in Figure 3b, the $F_{3\text{-order}}+$ has the greatest ability to resolve two adjacent emitters with FWHM at 130 nm compared to the initial value of $F_1$ at 190 nm, with an improvement factor of 1.4. The discrepancy in the resolution enhancement between the simulation (1.4) and the theoretical values ($\sqrt{3}$) is due to noise and the third-order polynomial function being only an approximation of the entire Taylor series. Additionally, we simulate the X-bar design labeled with emitters as continuing crossline structures (Figure S5). The optical images in Figure 3c-h are generated using the same background and excitation conditions as in Figure 3a. For the X-bar with emitters, the $F_{3\text{-order}}+$ (Figure 3h) accurately locates the lines down to 130 nm in position and exhibits the contour profile, demonstrating its capacity in optical nanoscopy for resolving line structures. Although the generated image's signal-to-noise ratio degrades as the number of processes rises,[37] we employ only three-order processes in this work, thus the signal-to-noise ratio remains within an acceptable range.

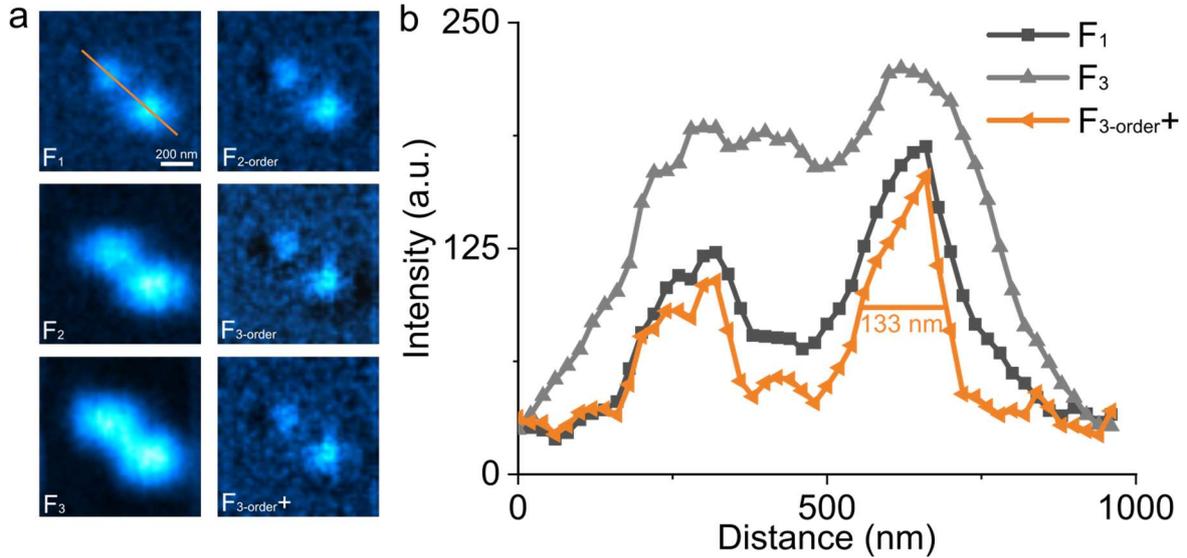

**Figure 4. Experimental demonstration of the sub-diffraction-limit imaging.** (a) The 455 nm emission band image of UCNPs under a 980 nm excitation (0.1 mW, 0.2 mW, 0.3 mW) and their corresponding results with a weighted finite difference method. (b) Line profiles of two nearby UCNPs from (a). Pixel dwell time: 1 ms. Pixel size: 20 nm. The scale bar is 200 nm.

To verify the resolving capability of the differential excitation techniques, we apply this approach to image single UCNPs in a sub-diffraction area (Figure 4a). The UCNPs are monodispersed randomly on a cover glass. We demonstrate the ability to resolve the discrete nanoparticles with a discernible distance of 350 nm. The lower excitation power (0.1 mW) enables imaging with higher resolutions than larger excitation powers (0.2 mW and 0.3 mW), which is consistent with our simulation results. In the highly doped ions system, the pumping rate at the excited energy level responds nonlinearly to variations in excitation power. Accordingly, the carrier distribution in an intermediate state can be controlled by modulating the excitation power. We find that decreasing the excitation power decreases the carrier population generation rate, allowing for a steeper nonlinear optical response and consequently shrinking the PSF. However, the signal-to-noise ratio of the obtained images degrades when the excitation power decreases. As shown in Figure 4a, the signal-to-noise ratio drops from 11.1 to 3.1 when changing the excitation power from 0.3 mW to 0.1 mW. Although it is theoretically possible to further compress the PSF by decreasing the excitation power, it is hard to achieve this experimentally owing to the deteriorating signal-to-noise ratio. Our proposed dynamic nonlinearity technique is a way to improve the resolution without further reducing excitation power. The line profiles of the processed image of a single nanoparticle (Figure 4b) further demonstrate the quantified result of the significantly reduced FWHM of 133 nm from 220 nm.

**Conclusions**

In conclusion, we report steep and dynamic nonlinearity in UCNPs under gradient excitation intensities, realizing super-resolution imaging with a confocal microscope. To the best of our knowledge, this is the first work using dynamic optical nonlinear responses within the superlinear region. The idea behind this work is to extract higher spatial-frequency information from dynamic nonlinearity to improve resolution in the superlinear technique. Unlike techniques that require the entire nonlinear curve of all the detected pixels,[39–41] our method shows that two or three images with nonlinear variations are sufficient to offer a substantial gain in resolution.[37,42] Compared to previous techniques that require high excitation intensities to reach adequate saturation, our proposed approach is achievable under ultra-low excitation power. Furthermore, this method has the potential to improve axial resolution without photobleaching or deterioration issues.[8]

This work employs highly dual-doped ion ($Yb^{3+}$ and $Tm^{3+}$) nanoparticles as fluorescent probes to produce a nonlinear dependency of four-photon 455 nm emission intensity on excitation power under a continuous-wave laser. The nonlinear slope can be increased by optimizing the activator/sensitizer concentration or designing a core-shell structure, which shifts the emission curve towards lower excitation powers without affecting optical properties.[15] This approach may also be used for emission bands from other excited energy levels due to the multiple long-lived intermediates in UCNPs. For example, compared to emissions from higher energy levels, the 800 nm emission band from the lower energy level has a lower threshold, which could be of benefit in deep tissue super-resolution imaging at a low-power density.[43–45] Although the demonstrated resolution is only a moderate improvement compared to conventional super-resolution techniques, like STED and STORM, our proposed method holds the potential to produce a narrower PSF if we can improve the signal-to-noise ratio beyond our results in this work. Our technique can also utilize the large nonlinear response induced by the photo-avalanching effect[35,36] in UCNPs to enhance resolution, though the imaging speed would be limited by the milliseconds of photon avalanche building time. Moreover, by taking advantage of heterochromatic nonlinear responses,[46] we can circumvent multiple image acquisition procedures under differential excitation powers. It will be fascinating to verify this approach in future work to simultaneously produce a series of multi-color images at differential saturating degrees via parallel detection channels. We anticipate that this work will provide new insights into dynamic nonlinear fluorescent probes, expanding applications in super-resolution imaging.

**Experimental Section**
The experimental details are provided in the Supporting Information

**Supporting Information**
Supporting information is available from the author.

**Acknowledgments**
The authors acknowledge the financial support from the China Postdoctoral Science Foundation (2021M702581), Guangdong Basic and Applied Basic Research Fund (2021A1515111016), Australian Research Council DECRA Fellowship (DE200100074-F.W.), and Dongguan Science and Technology Commissioner Project (20201800500172).

**Conflict of Interest**
The authors declare no conflict of interest.

# Exploiting dynamic nonlinearity in upconversion nanoparticles for super-resolution imaging

Chaohao Chen[1,2†], Lei Ding[3†], Baolei Liu[4], Ziqin Du[3], Yongtao Liu[5], Xiangjun Di[3], Xuchen Shan[4], Chenxiao Lin[6], Min Zhang[1], Xiaoxue Xu[7], Xiaolan Zhong[4], Jianfeng Wang[4], Lingqian Chang[8], Ben J. Halkon[9], Xin Chen[2], Faliang Cheng[1*], Fan Wang[4,10*]

[1]Guangdong Engineering and Technology Research Center for Advanced Nanomaterials, School of Environment and Civil Engineering, Dongguan University of Technology, Dongguan 523808, China

[2]Department of Chemical Engineering, Shaanxi Key Laboratory of Energy Chemical Process Intensification, Institute of Polymer Science in Chemical Engineering, School of Chemical Engineering and Technology, Xi'an Jiaotong University, Xi'an 710049, China

[3]School of Mathematical and Physical Sciences, Faculty of Science, University of Technology Sydney, NSW 2007, Australia

[4]School of Physics, Beihang University, Beijing 100191, China

[5]Smart Computational Imaging Laboratory, Nanjing University of Science and Technology, Nanjing, Jiangsu Province 210094, China

[6]Department for Electrochemical Energy Storage, Helmholtz-Zentrum Berlin für Materialien und Energie, Hahn-Meitner-Platz, Berlin 14109, Germany

[7]School of Biomedical Engineering, Faculty of Engineering and Information Technology, University of Technology Sydney, NSW 2007, Australia

[8]Beijing Advanced Innovation Center for Biomedical Engineering, School of Biological Science and Medical Engineering, Beihang University, Beijing 100191, China

[9]Centre for Audio, Acoustics & Vibration, Faculty of Engineering & IT, University of Technology Sydney, Ultimo, NSW 2007, Australia

[10]School of Electrical and Data Engineering, Faculty of Engineering and Information Technology, University of Technology Sydney, NSW 2007, Australia.

[†]These authors contributed equally: Chaohao Chen, Lei Ding

[*]Correspondence to fan.wang@uts.edu.au, chengfl@dgut.edu.cn.

*Synthesis of nanoparticles*

The $Yb^{3+}$-$Tm^{3+}$-$NaYF_4$ UCNPs were synthesized using a co-precipitation method[1]. In a typical process, a 50 mL three-neck round-bottom flask was filled with a methanol solution (1 mmol) of $YCl_3 \cdot 6H_2O$, $YbCl_3 \cdot 6H_2O$ and $TmCl_3 \cdot 6H_2O$, together with oleic acid (6 mL) and 1-octadecene (15 mL). The resultant mixture was heated for 30 minutes at 150 °C to create lanthanide oleate complexes. The solution was brought to room temperature. Following that, a methanol solution (6 mL) containing NaOH (2.5 mmol, 100 mg) and $NH_4F$ (4 mmol, 148 mg) was added and stirred at 50 °C for 30 minutes before being gently heated to 150 °C and held under argon flow for 20 minutes to remove methanol and residual water. The solution was then rapidly heated at 300 °C under an argon flow for 1.5 hours before cooling to ambient temperature. The resultant core nanoparticles were precipitated by adding ethanol, collected by centrifugation at 9000 rpm for 5 minutes, and the final $NaYF_4$: $Yb^{3+}$, $Tm^{3+}$ nanocrystals were redispersed in cyclohexane at a concentration of 5 mg/mL after repeated washes with cyclohexane/ethanol/methanol.

*Measurements and characterizations*

The morphology of nanoparticles was determined using transmission electron microscopy (TEM) imaging (Philips CM10 TEM and Olympus Sis Megaview G2 Digital Camera) at a 100 kV operating voltage. A drop of a dilute suspension of nanocrystals was placed onto copper grids to prepare the samples. The TEM image is shown in Figure S1.

*Sample preparation*

A coverslip was cleaned with 100% ethanol and air-dried to prepare a sample slide. On the coverslip, 5 μl UCNPs suspension (diluted to 0.01 mg/ml in cyclohexane) was dropped. After air drying, a clean glass slide was used to cover the coverslip.

*Optical setup*

All measurements were made using custom-built microscope equipment equipped with a three-axis closed-loop piezo stage (stage body MAX311D/M, piezo controller BPC303; Thorlabs), as shown in Figure S2. A polarization-maintaining single-mode fiber-coupled 980 nm diode laser excites UCNPs (BL976-PAG900, controller CLD1015, Thorlabs). Electronically rotating the half-wave plate (HWP, WPH05M-980, Thorlabs) and a polarizing beam splitter (PBS, CCM1-PBS252/M, Thorlabs) can accurately adjust the excitation power. After collimation, the excitation beam is reflected using a short-pass dichroic mirror (DM, T875spxrxt-UF1, Chroma) and focused onto the sample slide using a high numerical aperture objective (UPlanSApo, 100/1.40 oil, Olympus). The same objective collects photoluminescence and separates it from the excitation beams using DM. The band-pass filter (BPF, FF01-448/20-25, Semrock) is used to purify the emission signals. The emission may be coupled into multimode fiber (MMF, M42L02, Thorlabs), where it can be detected using a single-photon counting avalanche photodiode (SPAD, SPCM-AQRH-14-FC, Excelitas). All powers are measured at the objective lens's rear aperture. Pixel dwell times are set to 1 millisecond.

*Upconversion energy transfer process*

To study the nonlinear optical response of the emission and energy transfer process in $Yb^{3+}$-$Tm^{3+}$ doped UCNP systems, we simplify the energy level diagram as shown in Figure S3. It is composed of two energy levels corresponding to the sensitizer $Yb^{3+}$ and six energy levels corresponding to the activator $Tm^{3+}$. $^2F_{7/2}$ and $^2F_{5/2}$, $^3H_6$, $^3H_5/^3F_4$, $^3F_{2,3}/^3H_4$, $^1G_4$, and $^1D_2$ represented as $s_1$, $s_2$, 1, 2, 3, 4, 5 respectively. $n_i$ is the population of photons on the energy level. $c_i$ (i=1,2,3,4) is the energy transfer ratio between the excited level of $Yb^{3+}$ on the ground and the intermediate levels of $Tm^{3+}$. $K_{ij}$ is the cross-relaxation coefficient between the state $i$ and $j$. $a_{ij}$ is the branching ratio from energy level $i$ to $j$. $W_i$ is the intrinsic decay rate of $Tm^{3+}$ on level $i$. $P_{980}$ is the pumping rate of $Yb^{3+}$.

$$\frac{dn_1}{dt} = -c_1 n_1 n_{S2} + a_{21} W_1 n_2 + a_{31} W_2 n_3 + a_{41} W_3 n_4 + a_{51} W_4 n_5 - k_{31} n_1 n_3 - k_{41} n_1 n_4 - k_{51} n_1 n_5 \ldots\ldots(1)$$

$$\frac{dn_2}{dt} = c_1 n_1 n_{S2} - c_2 n_2 n_{S2} - a_{21} W_1 n_2 + a_{32} W_2 n_3 + a_{42} W_3 n_4 + a_{52} W_4 n_5 + k_{41} n_1 n_4 + 2 k_{31} n_1 n_3 \ldots\ldots(2)$$

$$\frac{dn_3}{dt} = c_2 n_2 n_{S2} - c_3 n_3 n_{S2} - (a_{31} + a_{32}) W_2 n_3 + a_{43} W_3 n_4 + a_{53} W_4 n_5 + 2 k_{51} n_5 n_1 + k_{41} n_4 n_1 - k_{31} n_3 n_1 \ldots\ldots(3)$$

$$\frac{dn_4}{dt} = c_3 n_3 n_{S2} - c_4 n_4 n_{S2} - (a_{43} + a_{42} + a_{41}) W_3 n_4 + a_{54} W_4 n_5 - k_{41} n_1 n_4 \ldots\ldots(4)$$

$$\frac{dn_5}{dt} = c_4 n_4 n_{S2} - (a_{54} + a_{53} + a_{52} + a_{51}) W_4 n_5 - k_{51} n_1 n_5 \ldots\ldots(5)$$

$$\frac{dn_{S2}}{dt} = P_{980} n_{S1} - w_S n_{S2} - (c_1 n_1 + c_2 n_2 + c_3 n_3 + c_4 n_4) n_{S2} \ldots\ldots(6)$$

To comprehend the slope of the emission curve in the superlinear region, five energy level rate equations with a nonlinear response excitation process were constructed. The parameters utilized in the full rate equations are listed in Table S1. Due to the greater obsecration cross-section of $Yb^{3+}$ at 980 nm, the excitation photons are absorbed by $Yb^{3+}$ solely and subsequently transferred to $Tm^{3+}$ during the excitation process. According to the established rate equations, the ground state has the highest population of carriers at the beginning. By transferring the pumping energy from $Yb^{3+}$ to the ground state of $Tm^{3+}$, the new distribution of the carriers will start from the ground state to the excited states of $Tm^{3+}$ in order of energy levels. The photon upconversion process is a sequential absorption process, with photon energy linearly absorbed by $Yb^{3+}$ in the UCNPs. The activator ions have multiple long-lived real intermediate states to facilitate multiphoton upconversion emission. The model displays the 455nm power dependency curve given in Figure S4, indicating optical response behavior in agreement with the experiments.

Although the overall 455 nm emission band involves various energy redistribution mechanisms, we find that the emission at low excitation irradiance is similar to a conventional multiphoton phenomenon, enabling Taylor series analysis in this work.[2]

*Generation of simulated data*

This section describes the underlying algorithms and performs theoretical simulations to assess the performance in nanoscopy. The details of the system and the imaging procedure for the point-scanning microscopy have been described in our recent work[3]. Briefly, the emission is collected by an objective lens with a high numerical aperture (NA = 1.4) and focused by the tube lens onto the single-photon detector, so that the effective point spread function (PSF) ($h_{em}(x,y)$) can be described as:

$$\begin{cases} h_{eff}(x,y) = h_{em}(x,y) \times h_c(x,y) \\ h_{em}(x,y) = \eta(i) \times h_{exc}(x,y) \end{cases} \quad (7)$$

Here $h_{em}(x,y)$ is the PSF of emission; $h_c(x,y)$ is the PSF of the confocal collection system; $h_{exc}(x,y)$ is the PSF of the excitation beam (Gaussian beam); and $\eta(i)$ is the excitation power dependent emission intensity curve. The full width at half-maximum (FWHM) of the intensity in PSF ($h_{exp}(x,y)$) represents the nanoscopy resolution. The experimentally measured intensity distribution PSF ($h_{exp}(x,y)$) on the image plane in our system is the convolution between $h_{eff}(x,y)$ and the spatial distribution profile ($h_{UCNP}(x,y)$) of the nanoparticle as below:

$$h_{exp}(x,y) = h_{eff}(x,y) * h_{UCNP}(x,y) \qquad (8)$$

*Imaging algorithm*

The idea behind this work is to extract higher spatial-frequency information from nonlinear optical responses of upconversion nanoparticles to further improve the resolution achieved in the superlinear technique. Inspired by photobleaching imprinting microscopy[4] and stepwise optical saturation microscopy[2], we find that dynamic variables in the PSF can be leveraged to extract higher-order spatial-frequency information.

Following the simulation results and a similarly reported derivation[2], we can express the fluorescence intensities of the i-th step $F_i(x)$ among the M images, as a sum of linear and nonlinear components.

$$F_i(x) = KN_0(aI_{0i}^N g^N(x) - a^2 I_{0i}^{2N} g^{2N}(x) + a^3 I_{0i}^{3N} g^{3N}(x) - \cdots) \qquad (9)$$

Where $K = \psi_F/\tau$, $\psi_F$ is the fluorescence detection efficiency, $\tau$ is the fluorescence lifetime, $N_0$ is the concentration of the fluorophore, $a = \tau\sigma_N(\lambda/hc)^N$, $N$ is N-photon excitation, $\sigma_N$ is the cross-section for N-photon excitation, $\lambda$ is the excitation wavelength, $h$ is Planck's constant, $c$ is the velocity of light, $I_{0i}$ is the focal irradiance of the i-th step, $g(x) = \exp(-2x^2/\omega_0^2)$. In Equation (9), the higher power of $g^N(x)$, such as $g^{2N}(x)$, $g^{3N}(x)$, indicates components of the higher spatial information, whereas $g(x)$ represents components of diffraction-limited information. With the definition of $g(x)$, we can have a $\sqrt{N}$ improvement of the resolution for N-photon excitation.[7] Thus, an M-th powers of N-photon excitation, $g^{MN}(x)$, produces a $\sqrt{MN}$-fold increase in spatial resolution.

Consequently, the strategy is to eliminate the lowest M-1 powers of $g^N(x)$. The lowest power of $g^N(x)$ dominates the spatial resolution of $F_i(x)$, according to the power series difference among components in Equation (9). The resulting image is $F_{M-ord}(x) = \sum_{i=i}^{M} c_i F_i(x)$, where the coefficients $c_i$ are chosen such that the lowest power of $g^N(x)$ in $F_{M-ord}(x)$ is $g^{MN}(x)$. Table S2 presents the coefficients $c_i$ for the weighted sum of the algorithm from two-order to three-order. For example, for extracting 2-order high spatial-frequency information, we obtain two images $F_1(x)$ and $F_2(x)$ under different excitation power $I_1$ and $I_2$, respectively. We can process two images with 2-order weighted finite difference method:

$$F_{2-o}(x) = KN_0[-a^2 I_1^N(I_1^N - I_2^N)g^{2N}(x) + a^3 I_1^N(I_1^{2N} - I_2^{2N})g^{3N}(x) - \ldots] \quad (10)$$

As the resolution is dominated by $g^{2N}(x)$, $F_{2-order}(x)$ yields a super-resolution image with $\sqrt{2}$-fold increase in spatial resolution. Similarly, three images has $\sqrt{3}$-fold enhancement with 3-order weighted finite difference process:

$$F_{3-order}(x) = KN_0 \left[ a^3 \left( I_1^{3N} - \frac{I_2^{2N} I_1^N (I_1^N - I_3^N) + I_3^{2N} I_1^N (I_1^N - I_2^N)}{I_2^N - I_3^N} \right) g^{3N}(x) - \ldots \right] \quad (11)$$

We apply Fourier spectrum modulation[6] to further increase the signal-to-noise ratio and resolution. The Fourier modulation process is summarized in Equation (12). For the imaging procedure, we choose $F_1(x)$ and $F_{i-order}(x)$. Firstly, we subtract $F_{i-order}(x)$ from $F_1(x)$ to produce $F_1'(x)$. Secondly, we use the fast Fourier transform (FFT) technique to convert both images of $F_{i-ord}(x)$ and $F_1'(x)$ to the frequency domain. Thirdly, we conduct the operation by subtracting the frequency spectrum of $F_1'(x)$ from $F_{i-order}(x)$, which helps to remove low-frequency information and background noise. Finally, applying an inverse fast Fourier transform (IFFT) to the removed frequency spectrum generates an image with an increased signal-to-noise ratio and resolution.

$$F_{i-order}(x) += IFFT(r * FFT(F_{i-ord}) - FFT(F_1 - F_{i-order})) \quad (12)$$

Where r is the weighting factor.

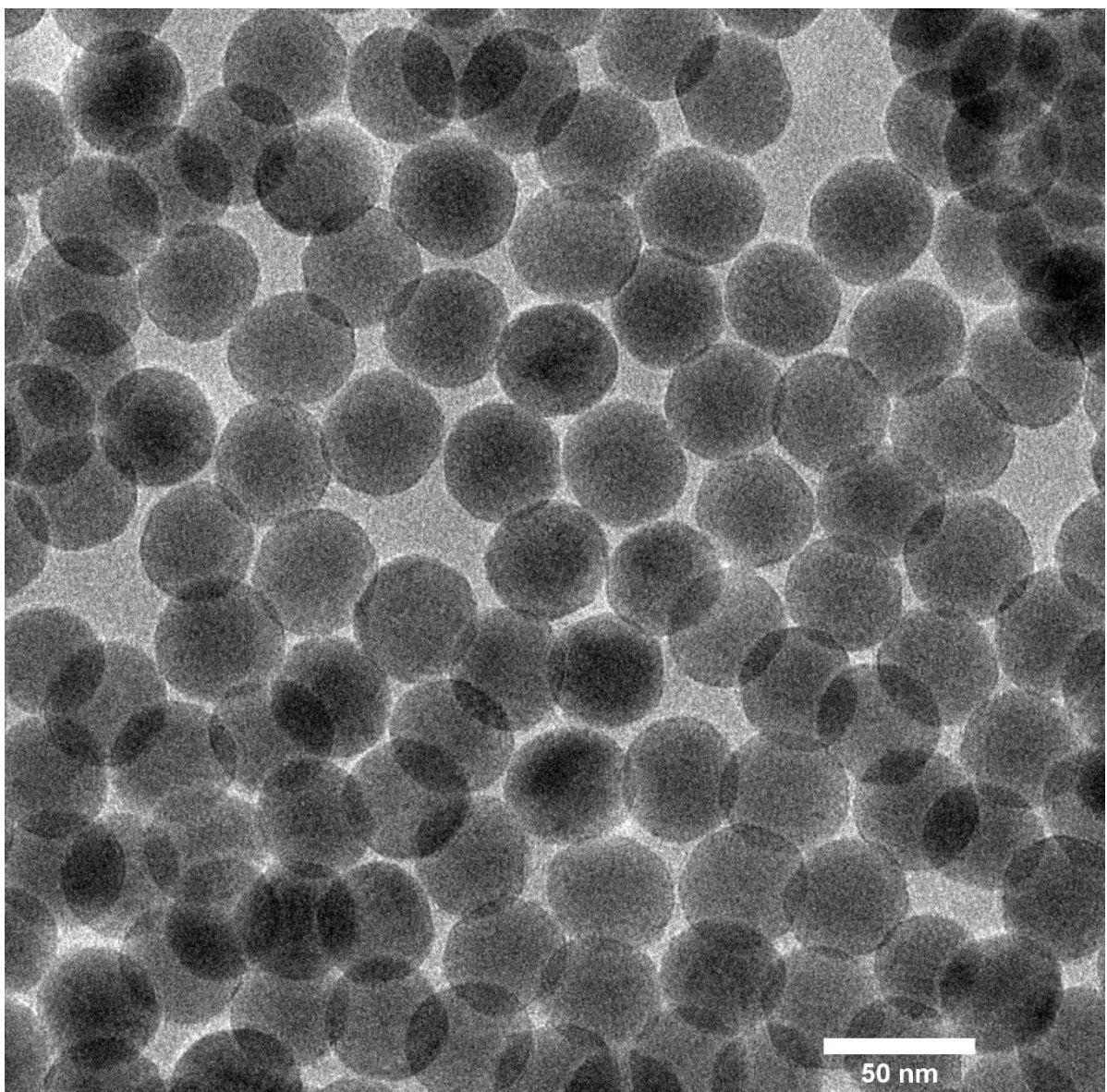

Figure S1. TEM image of 41 nm single nanoparticle (NaYF$_4$: 40% Yb$^{3+}$, 4% Tm$^{3+}$).

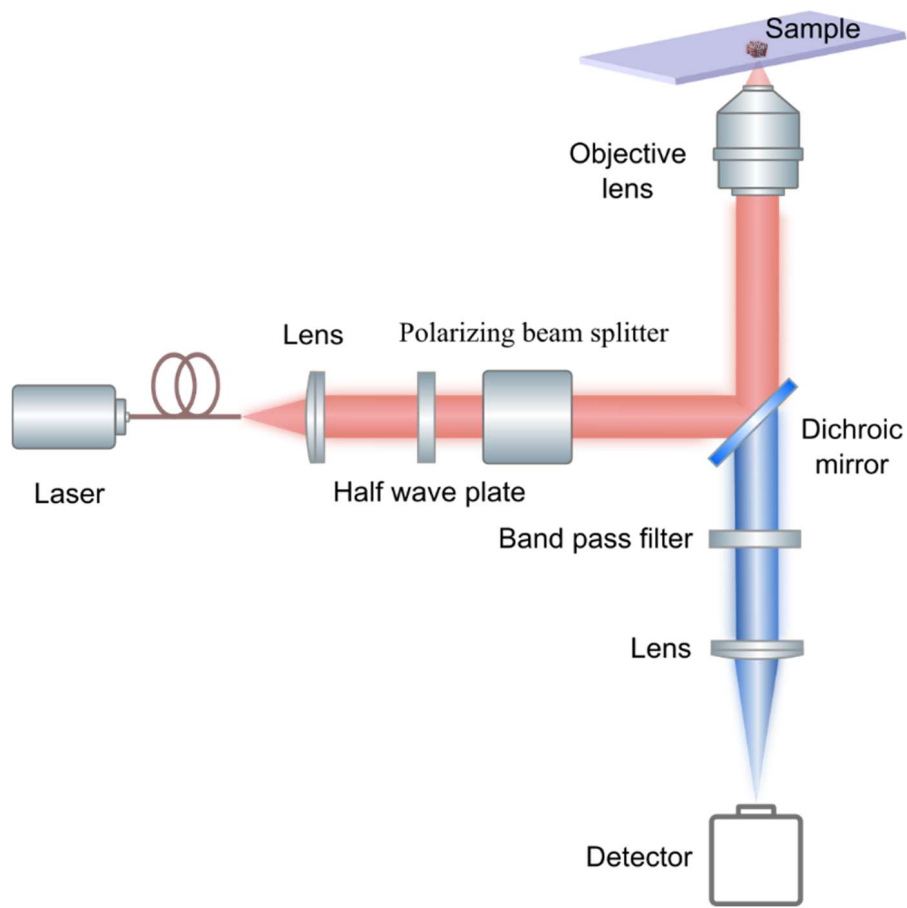

Figure S2. Schematic setup of the single beam scanning super-resolution nanoscopy

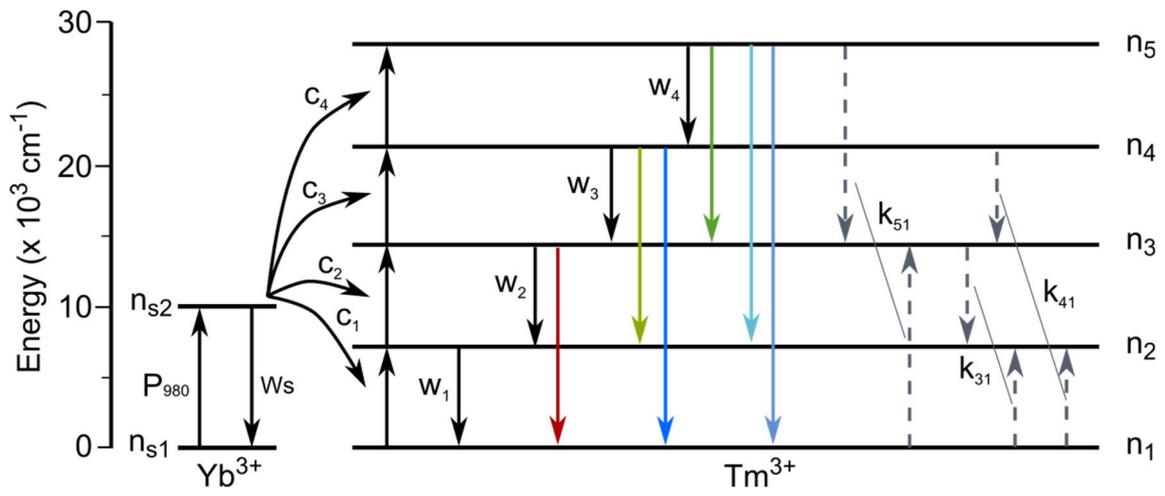

Figure S3. The simplified energy level diagram of $Tm^{3+}$ and $Yb^{3+}$ doped UCNP.

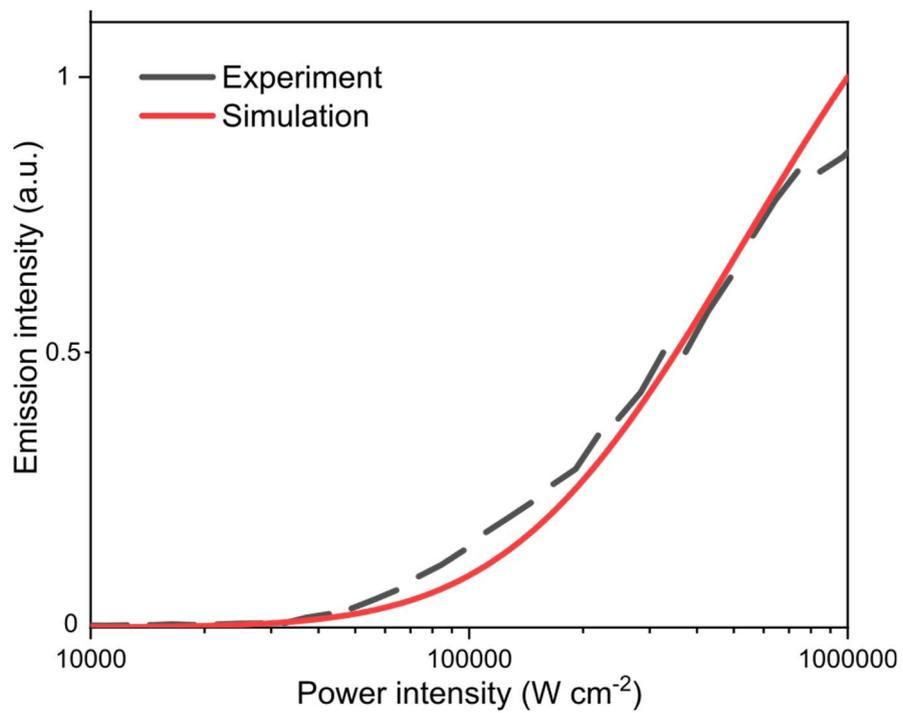

Figure S4. The simulated and experimental emission intensity curve of the 455 nm emissions from $Tm^{3+}$ and $Yb^{3+}$ doped UCNP under 980 nm excitation.

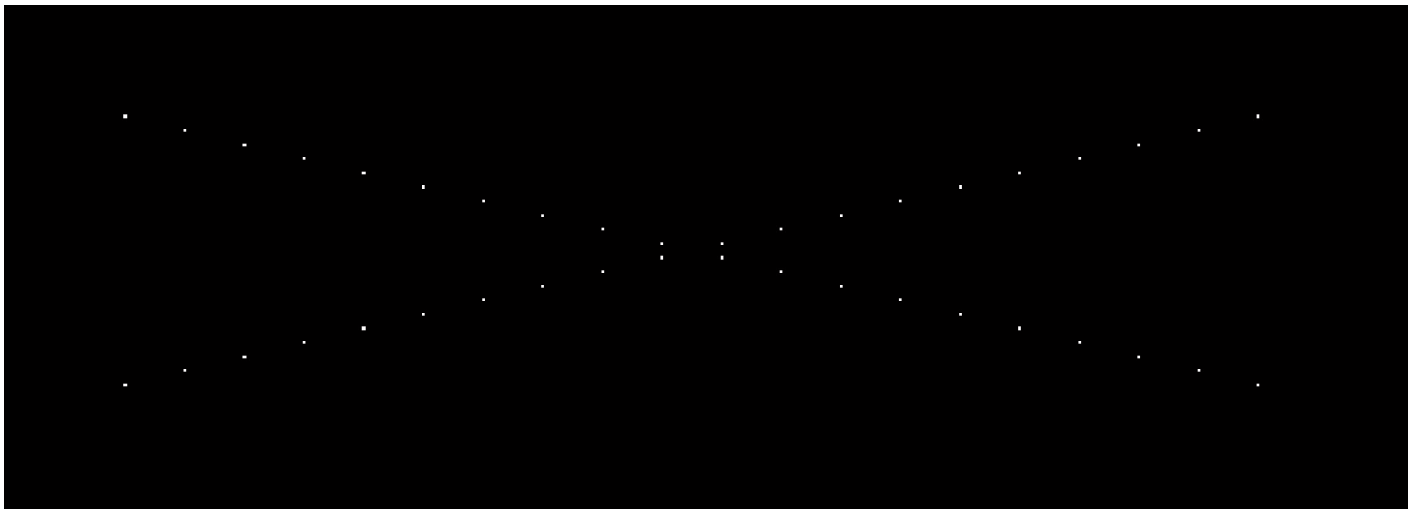
Figure S5. The imaging object: X-bar design labeled with emitters as continuing crossline structures.

**Table S1 Parameters of the internal transition and spontaneous radiation probability in Tm and Yb**

| $w_4$ (s$^{-1}$) | $w_3$ (s$^{-1}$) | $w_2$ (s$^{-1}$) | $w_1$ (s$^{-1}$) | $w_s$ (s$^{-1}$) |
|---|---|---|---|---|
| $3.2\times10^3$ | $1.4\times10^4$ | $1.8\times10^4$ | $6.4\times10^3$ | $7.6\times10^3$ |
| Branching ratio from energy level i to j | | | | |
| $a_{51}$ | $a_{52}$ | $a_{53}$ | $a_{54}$ | $a_{21}$ |
| 0.24 | 0.23 | 0.2 | 0.33 | 0 |
| $a_{41}$ | $a_{42}$ | $a_{43}$ | $a_{31}$ | $a_{32}$ |
| 0.18 | 0.24 | 0.58 | 0.27 | 0.73 |
| Energy transfer rates | | | | |
| $c_1$ (s$^{-1}$) | $c_2$ (s$^{-1}$) | $c_3$ (s$^{-1}$) | $c_4$ (s$^{-1}$) | |
| $6.3\times10^4$ | $5\times10^4$ | $7\times10^4$ | $5\times10^3$ | |
| Cross-relaxation coefficients | | | | |
| $k_{51}$ (s$^{-1}$) | $k_{41}$ (s$^{-1}$) | $k_{31}$ (s$^{-1}$) | $P_{980}$ (s$^{-1}$) | |
| $4.8\times10^5$ | $1.75\times10^5$ | $1.5\times10^5$ | $2.8\times10^5$ | |

**Table S2 Coefficients relating to the weighted finite difference method**

| Coefficients | | Expressions |
|---|---|---|
| **2-order** | $c_1$ | $1$ |
| | $c_2$ | $-\dfrac{I_1^N}{I_2^N}$ |
| **3-order** | $c_1$ | $1$ |
| | $c_2$ | $-\dfrac{I_1^N(I_1^N - I_3^N)}{I_2^N(I_2^N - I_3^N)}$ |
| | $c_3$ | $-\dfrac{I_1^N(I_1^N - I_2^N)}{I_3^N(I_2^N - I_3^N)}$ |